\documentclass[11pt,a4paper,jhep]{article}
\pdfoutput=1
\usepackage{jheppub}
\usepackage{textcomp}
\usepackage{kotex}
\usepackage{tikz}
\usepackage{cancel}
\usepackage{verbatim}
\usepackage{float}
\usepackage{graphicx,subfigure,color,array,slashed,wasysym}
\usepackage{caption}
\usepackage{dcolumn}
\usepackage{bm}
\usepackage{amsmath,amssymb,amsbsy,amsfonts,latexsym,graphicx}
\usepackage{bbold}

      \renewcommand{\t}{\tau}

\newcommand{\be}{\begin{equation}}
\newcommand{\ee}{\end{equation}} 
\newcommand{\ba}{\begin{array}}
\newcommand{\ea}{\end{array}}
\newcommand{\bea}{\begin{eqnarray}}
\newcommand{\eea}{\end{eqnarray}}

\newcommand{\half}{\frac{1}{2}}

\renewcommand{\Im}{{ \rm Im}}

\usepackage{multirow}

\newcommand\hexa[2]{
	\draw[line width= 2pt] (#1-0.6,#2-0.5)--(#1+0.4,#2-0.5);
	\draw[line width= 2pt] (#1-0.4,#2+0.5)--(#1+0.6,#2+0.5);
	\draw[line width= 2pt] (#1-0.4,#2+0.5)--(#1-1,#2);
	\draw[line width= 2pt] (#1-0.6,#2-0.5)--(#1-1,#2);
	\draw[line width= 2pt] (#1+0.6,#2+0.5)--(#1+1,#2);
	\draw[line width= 2pt] (#1+0.4,#2-0.5)--(#1+1,#2);
	\draw [fill=white](#1-1,#2) circle (0.1);
	\draw [fill=black](#1+1,#2) circle (0.1);
	\draw [fill=black](#1-0.4,#2+0.5) circle (0.1);
	\draw [fill=black](#1-0.6,#2-0.5) circle (0.1);
	\draw [fill=white](#1+0.6,#2+0.5) circle (0.1);
	\draw [fill=white](#1+0.4,#2-0.5) circle (0.1);
}

\tikzset{%
	ebo unit/.store in=\ebounit,
	ebo corners/.style={rounded corners=#1\ebounit},
}

\title{ \Large  ABC-stacked 
	multilayer graphene in holography} 
\author[a]{Jeong-Won Seo,}
\author[a]{Taewon Yuk,}
\author[a]{Young-Kwon Han,}
\author[a]{Sang-Jin Sin}
\emailAdd{1113dino@naver.com}
\emailAdd{tae1yuk@gmail.com}
\emailAdd{youngkwonhan346@gmail.com}
\emailAdd{sjsin@hanyang.ac.kr}
\affiliation[a]{ Department of Physics, Hanyang University, Seoul 04763, Korea }
\abstract{  
	A flat band can be studied an infinitely strong coupling, realized in a simple system. Therefore, its holographic realization should be interesting. Laia and Tong gave a realization of the flat band over the entire momentum region by introducing a particular boundary term.
	Here, we give a model with a flat band over a finite region of momentum space using a bulk interaction term instead of the boundary term. We find that the spectrum of our model is precisely analogous to that of the ABC stacked multilayer graphene. In the presence of the chemical potential, the flat band is bent in our holographic model, which is very close to the band deformation due to the spin-orbit 
}

\subheader{\today }
\arxivnumber{2007.xxxxx} 
\begin{document}
	\maketitle
	
	\section{Introduction} 
	
	Flat band systems have been widely studied to realize a strongly interacting system in relatively simple systems: the effective coupling, which is the ratio of the potential to kinetic energy, diverges in the flat band limit. 
	It is known to exist in artificial lattices with compact localized states \cite{Weeks:2010:10.1103_PhysRevB.82.085310, Guo:2009:10.1103_PhysRevB.80.113102, Goldman:2011:10.1103_PhysRevA.83.063601, Beugeling:2012:10.1103_PhysRevB.86.195129, Tadjine:2016:10.1103_PhysRevB.94.075441, Li:2016:10.1039_C6NR03223K}, and has been formed in optical lattices\cite{Wu:2007:10.1103_PhysRevLett.99.070401, Apaja:2010:10.1103_PhysRevA.82.041402, Shen:2010:10.1103_PhysRevB.81.041410, Guzman-Silva:2014:10.1088_1367-2630_16_6_063061, Mukherjee:2015:10.1103_PhysRevLett.114.245504, Vicencio:2015:10.1103_PhysRevLett.114.245503, Taie:2015:10.1126_sciadv.1500854, Xia:2016:10.1364_OL.41.001435, Diebel:2016:10.1103_PhysRevLett.116.183902}. 
	In the real material, however, it was realized only recently in the magic angle twisted bilayered graphene(MATBG)\cite{cao2018correlated, cao2018unconventional}.
	
	While mono-layer graphene has a simple band with Dirac cones in K points\cite{novoselov2005two,zhang2005experimental}, 
	multilayered graphene has rich band structures depending on the stacking order\cite{haering1958band,mcclure1969electron,PhysRevB.37.8327,PhysRevB.43.4579,PhysRevB.46.4540} or twisting angles\cite{Morell:2010:10.1103_PhysRevB.82.121407,marchenko2018extremely,shen2020correlated}. The electronic structure of multilayered graphene (MLG) is well described by a tight binding model\cite{PhysRevLett.97.266801,PhysRevB.73.245426,PhysRevB.73.214418,charlier1994first,PhysRevB.81.125304} : the so-called ABC stacked  multilayer graphene (ABC-MLG) is a MLG that has the stacking in figure \ref{fig:stackingstruc} and such   structure is observed often in nature, e.g., graphite\cite{bacon1950note, gasparoux1970etude}. Its band structure has a flat band over a finite region of the Brillouin zone \cite{PhysRevB.81.125304}, which is characteristically different from those appearing in the MATBG, where the flat band exists over the entire momentum region. This flat band over a finite region was also confirmed in the ARPES data\cite{PhysRevB.97.245421, pierucci2016atomic, wang2018flat}. 

	Since the flat band study was motivated by the strong correlation, it would be interesting to study it using the gauge gravity duality\cite{Maldacena:1997re, Witten:1998qj} and ask what would be the effect of strong interaction there. 
	Laia and Tong\cite{laia2011holographic} discovered a holographic model that realizes a flat band by introducing the boundary term. 
	$$\int_{bdy} {\bar\psi}\Gamma^{xy}\psi,$$ for the Dirac spinor. 
	Recently, we pointed out that their model has not only a flat band but also a Dirac band, which is relevant to the stabilized Lieb lattice\cite{Han:2022pjz}.
	
	In this paper, we study another holographic model with a particular bulk interaction term $$\int_{bulk}B_{xy} {\bar\psi}\Gamma^{xy}\psi,$$ instead of the particular boundary term. According to \cite{oh2021ginzberg}, such Yukawa coupling introduces special features to the fermion spectrum depending on the broken symmetry. In our case, the spectral function has a flat band over a disk due to the broken symmetry, which is the isotropic scaling in $x,y$ directions. Since $\Gamma^{xy}$ happens to be the generator of the rotational symmetry of the x-y plane, there is rotational symmetry in the spectrum. 
	We will see that the flat band in our holographic bulk is overwhelmingly similar to that of the ABC stacked graphene system, and there is a simple relationship between the 
	holographic parameter and the tight-binding model parameter.
	
	The most pressing question is what the holography predicts for the outcome of the strong correlation. We examined how the flat band is deformed under the chemical potential to see such an effect. 
	In non-interacting cases, the role of the chemical potential on the fermion spectrum is just shifting the energy level. In holographic theory, we find that, apart from the level shift, the disk-like flat band is bent like a bowl, which is characteristically different from the non-interacting theory. In the case of two flavors with different signs of $B_{xy}$, its spectral weight is similar to the effect of the level repulsion in the presence of the spin polarization. This suggests that the flavor index can be treated as a component index of the spin.
	
	\section{Tight-Binding Model of ABC-stacked multilayer graphene} 
	
	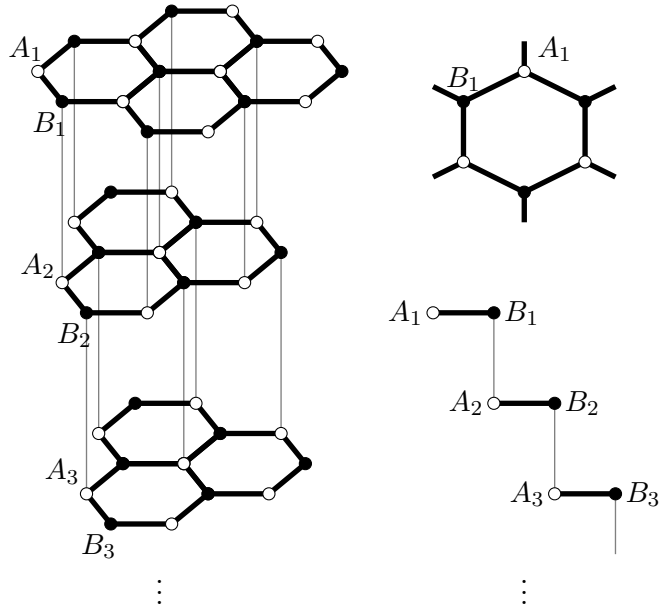
\begin{figure}[H]
		\centering
		\begin{tikzpicture}[scale=0.8]
			\draw[line width=0.5pt,color=gray] (-4.4, 4.5) -- (-4.4, 1.5);
			\draw[line width=0.5pt,color=gray] (-4.6, 3.5) -- (-4.6, 0.5);
			\draw[line width=0.5pt,color=gray] (-3, 4) -- (-3, 1);
			\draw[line width=0.5pt,color=gray] (-2.8, 5) -- (-2.8, 2);
			\draw[line width=0.5pt,color=gray] (-3.2, 3) -- (-3.2, 0);
			\draw[line width=0.5pt,color=gray] (-1.4, 4.5) -- (-1.4, 1.5);
			\draw[line width=0.5pt,color=gray] (-1.6, 3.5) -- (-1.6, 0.5);
			\draw[line width=0.5pt,color=gray] (-4, 1) -- (-4, -2);
			\draw[line width=0.5pt,color=gray] (-4.2, 0) -- (-4.2, -3);
			\draw[line width=0.5pt,color=gray] (-2.4, 1.5) -- (-2.4, -1.5);
			\draw[line width=0.5pt,color=gray] (-2.6, 0.5) -- (-2.6, -2.5);
			\draw[line width=0.5pt,color=gray] (-1, 1) -- (-1, -2);
			\node(dots) at (-3,-4.5){\vdots};
			\node[above] at (-5.2,4) {\(A_1\)};
			\node[below] at (-4.8,3.5) {\(B_1\)};
			\node[above] at (-5,0.4) {\(A_2\)};
			\node[below] at (-4.4,0) {\(B_2\)};
			\node[above] at (-4.6,-3) {\(A_3\)};
			\node[below] at (-4,-3.5) {\(B_3\)};
			\hexa{-4}{4};
			\hexa{-2.4}{4.5};
			\hexa{-2.6}{3.5};
			\hexa{-1}{4};
			\hexa{-3.4}{1.5};
			\hexa{-3.6}{0.5};
			\hexa{-2}{1};
			\hexa{-3}{-2};
			\hexa{-3.2}{-3};
			\hexa{-1.6}{-2.5};
			\draw[line width=2pt,color=black] (3, 4) -- (2, 3.5);
			\draw[line width=2pt,color=black] (3, 4) -- (4, 3.5);
			\draw[line width=2pt,color=black] (2, 3.5) -- (2, 2.5);
			\draw[line width=2pt,color=black] (4, 3.5) -- (4, 2.5);
			\draw[line width=2pt,color=black] (3, 2) -- (2, 2.5);
			\draw[line width=2pt,color=black] (3, 2) -- (4, 2.5);
			
			\draw[line width=2pt,color=black] (3, 2) -- (3, 1.5);
			\draw[line width=2pt,color=black] (3, 4) -- (3, 4.5);
			\draw[line width=2pt,color=black] (2, 3.5) -- (1.5, 3.75);
			\draw[line width=2pt,color=black] (4, 3.5) -- (4.5, 3.75);
			\draw[line width=2pt,color=black] (1.5, 2.25) -- (2, 2.5);
			\draw[line width=2pt,color=black] (4.5, 2.25) -- (4, 2.5);
			
			\draw [fill=white](3,4) circle (0.1);
			\draw [fill=black](2,3.5) circle (0.1);
			\draw [fill=black](4,3.5) circle (0.1);
			\draw [fill=black](3,2) circle (0.1);
			\draw [fill=white](2,2.5) circle (0.1);
			\draw [fill=white](4,2.5) circle (0.1);
			
			\node[above] at (3.5,4) {\(A_1\)};
			\node[above] at (2,3.5) {\(B_1\)};
			
			\draw[line width=2pt,color=black] (1.5, 0) -- (2.5, 0);
			\draw[line width=0.5pt,color=gray] (2.5, 0) -- (2.5, -1.5);
			\draw[line width=2pt,color=black] (2.5, -1.5) -- (3.5, -1.5);
			\draw[line width=0.5pt,color=gray] (3.5, -1.5) -- (3.5, -3);
			\draw[line width=2pt,color=black] (3.5, -3) -- (4.5, -3);
			\draw[line width=0.5pt,color=gray] (4.5, -3) -- (4.5, -4);
			
			\draw [fill=white](1.5,0) circle (0.1);
			\draw [fill=black](2.5,0) circle (0.1);
			\draw [fill=white](2.5,-1.5) circle (0.1);
			\draw [fill=black](3.5,-1.5) circle (0.1);
			\draw [fill=white](3.5,-3) circle (0.1);
			\draw [fill=black](4.5,-3) circle (0.1);
			
			\node[left] at (1.5,0) {\(A_1\)};
			\node[right] at (2.5,0) {\(B_1\)};
			\node[left] at (2.5,-1.5) {\(A_2\)};
			\node[right] at (3.5,-1.5) {\(B_2\)};
			\node[left] at (3.5,-3) {\(A_3\)};
			\node[right] at (4.5,-3) {\(B_3\)};	
			\node(dots) at (3,-4.5){\vdots};	
		\end{tikzpicture}
		\caption{The atomic structure of ABC-MLG. The white dots are for the \(A\) sublattice, and the black dots are for the \(B\) sublattice. Lines denote bondings. Thick black lines are for nearest neighbor couplings \(t_{\parallel}\) within the intralayer, and thin gray lines are for couplings  \(t_{\perp}\)  of interlayer vertical bonds that couples \(B_{j}\) and \(A_{j+1}\) for \(j=1,2,\cdots, N-1\). we call this kind of stacking order ABC-stacking or rhombohedral stacking.}
		\label{fig:stackingstruc}
	\end{figure}
	
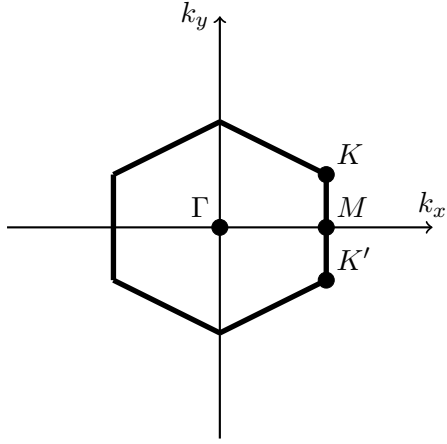
\begin{figure}[H]
		\centering
		\begin{tikzpicture}[scale=1.4]
			
			\draw[thick,->] (3,1) -- (3,5) node [left] {\(k_y\)};
			\draw[thick,->] (1,3) -- (5,3) node [above] {\(k_x\)};
			
			\draw[line width=2pt,color=black] (3, 4) -- (2, 3.5);
			\draw[line width=2pt,color=black] (3, 4) -- (4, 3.5);
			\draw[line width=2pt,color=black] (2, 3.5) -- (2, 2.5);
			\draw[line width=2pt,color=black] (4, 3.5) -- (4, 2.5);
			\draw[line width=2pt,color=black] (3, 2) -- (2, 2.5);
			\draw[line width=2pt,color=black] (3, 2) -- (4, 2.5); 
			\draw [fill=black](4,3.5) circle (0.075);
			\draw [fill=black](4,2.5) circle (0.075);
			\draw [fill=black](4,3) circle (0.075);
			\draw [fill=black](3,3) circle (0.075);
			
			\node[above right] at (4,3.5) {\(K\)};
			\node[above right] at (4,2.5) {\(K'\)};
			\node[above right] at (4,3) {\(M\)};
			
			\node[above left] at (3,3) {\(\Gamma\)};		
		\end{tikzpicture}
		\caption{Graphene honeycomb lattice in \(k\) space. The Dirac cones are located at the \(K\) and \(K'\) points, and we call them Dirac points.}
		\label{fig:Kpointstruc}
	\end{figure}

	The ABC-stacked multilayer graphene (ABC-MLG) has the structure given in the figure \ref{fig:stackingstruc}. Surface states are localized at the outermost layers of its band structure\cite{PhysRevB.81.125304, PhysRevB.73.245426, mcclure1969electron, lu2006absorption,PhysRevB.78.245416}.
	We can let full hamiltonian for the ABC-MLG system\cite{PhysRevB.81.125304} at the \(K\) point in Figure \ref{fig:Kpointstruc}. 
	\begin{align}
		\mathcal{H}=\begin{pmatrix} H_1 & V & & & \\ V^{\dagger} & H_2 & V & & \\ & V^{\dagger} & H_3 & V & & \\&  &  &\ddots & \end{pmatrix}, \quad \hbox{ where~~} 
		H_{j}=\begin{pmatrix} U_j & vp_{-} \\  vp_{+} & U_j \end{pmatrix}, && V=\begin{pmatrix} 0 & 0 \\  t_{\perp} & 0 \end{pmatrix} 
		\label{eq:fullH}
	\end{align}
	where  \(U_{j}\) is the electronic potential at \(j\)th layer, \(p_{\pm}=p_x\pm ip_y\) with \(\mathbf{p}=-i\nabla\) and \(v\) is the band velocity of monolayer given by \(v=\sqrt{3}a t_{\parallel}/2\hbar\).
	
	When \(U_j=0\), the hamiltonian gives eigenvalues as \cite{PhysRevB.81.125304}
	\begin{equation} 
		\varepsilon_{n}=\pm\sqrt{(vp)^2+t_{\perp}^2+2t_{\perp}vp\cos\varphi_n},
	\end{equation}
	where \(\varphi_n\) (\(n=1,2, \cdots, N\)) is a solution of the equation
	\begin{equation}
		vp\sin(N+1)\varphi_n+t_{\perp}\sin(N\varphi_n)=0.
		\label{eq:phi}
	\end{equation}
	By imposing the standing wave condition of the bulk wave function in the z-direction, that is, vanishing at the sites \(B_0\) and \(A_{N+1}\), we can get the corresponding wavefunction:
	\be
	\left| \psi \right\rangle =\psi(A_1) \left| A_1 \right\rangle + \psi(B_1) \left| B_1 \right\rangle + \cdots, \ee
	with 
	\begin{equation}
		\begin{pmatrix} \psi(A_j) \\  \psi(B_j)  \end{pmatrix}=C \begin{pmatrix} e^{i\theta(j-1)}\sin(N+1-j)\varphi_n \\  \pm e^{i\theta j}\sin j\varphi_n  \end{pmatrix},
	\end{equation}
	where \(\theta=\arctan (p_{y}/p_{x})\) and \(C\) is normalization factor. In the bulk limit, \(\varphi_n\) would become the wavenumber in z-direction along the layer stacking \cite{PhysRevB.81.125304}.  
	
	Suppose we draw the band structure of ABC-MLG from the full hamiltonian. In that case, we can see the zero energy band is the same as that of the effective hamiltonian given in Eq. (\ref{effspec}), and the size of the flat band saturates to a fixed value \(t_{\perp}\) in the large $N$ limit.
	
	\begin{figure}[H]
		\centering
		\subfigure[\(N=3\)]
		{\includegraphics[width=3.5cm]{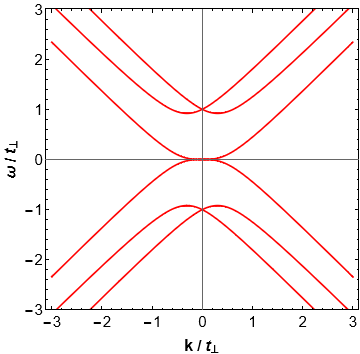}}
		\subfigure[\(N=10\)]
		{\includegraphics[width=3.5cm]{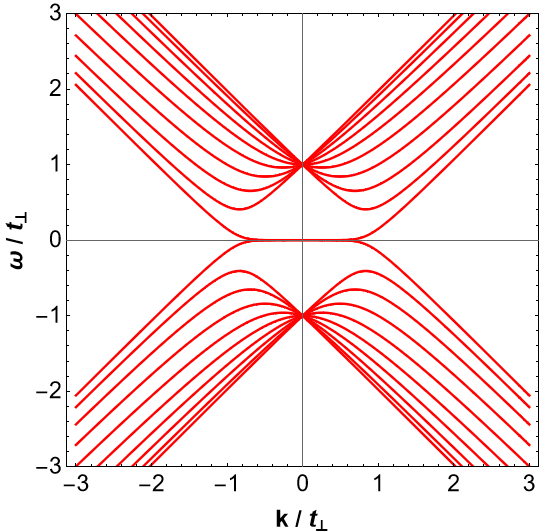}}
		\subfigure[\(N=20\)]
		{\includegraphics[width=3.5cm]{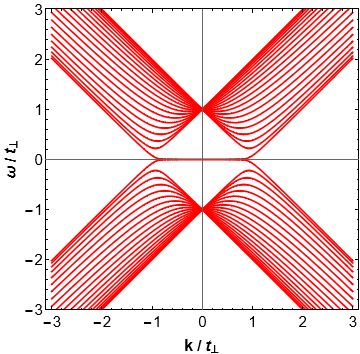}}
		\subfigure[ 3d overview]
		{\includegraphics[width=4cm]{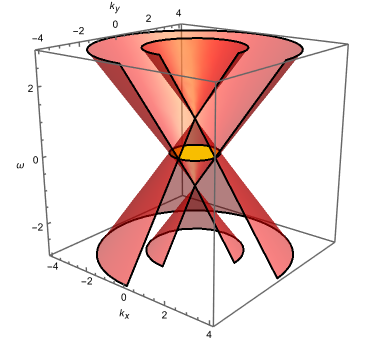}}
		\caption{Self-consistent band structure of ABC-MLG with  \(N\)  layer stacking. In (d), the orange disk at the center is the flat band of size \(R_{tb}\).   We used  \(t_{\perp}=1\) and \(U_{j}=0\).}
		\label{fig:AMGband}
	\end{figure}
	
	For large enough \(N\), the presence of the flat band between the two Dirac cones and the rotational symmetry around the \(\omega\) axis implies that this flat band is disk-shaped. The size of the flat band is the radius of the flat disk \(R_{tb}\).

	Defining flat band as the region where the band gap between the upper band and lower band is smaller than 0.01, we can calculate \(R\) with stacking number \(N\). Numerically we find 
	\be
	R_{tb}= t_{\perp}e^{-1/gN},
	\ee 
	with \(g\sim1/4\).
	It is interesting to notice the formal similarity of this relation with 
	$ 
	\Delta\sim e^{-\frac{1}{g^{2}N(0)}},  $
	of BCS superconductivity where $\Delta$ is the gap, $g$ is the electron-phonon coupling, and $N(0)$ is the density of state at the fermi surface. 
	In  the large $N$  limit, we  see that the size of the 
	flat band in the tight binding model is given by the interlayer coupling:
	\be
	R_{tb}= t_{\perp}.
	\label{eq:tbfbr}
	\ee	
	From the three-dimensional point of view, 
	the flat band of the ABC-MLG can be figured as the surface states localized at outer layers,  although it is a consequence of the superposition of a large number of the Graphene layers. The size of the flat band created here is related to the number of stacking and scaled by interlayer coupling \(t_{\perp}\) \cite{PhysRevB.73.245426}.
	
	\section{Holographic model of ABC-stacked multilayer graphenes}

	\subsection{Flow equation formula and Holographic model}
	The action is given by the sum $S=S_{ \psi}+S_{bdy}+S_{g,A,\Phi}+S^A_{int}+S^B_{int},$ where 	
	\begin{align} \label{eq:action1}
		& S_{ \psi} =i\int d^{4}x\; \sum_{j=1}^{2}\sqrt{-g}\bar{\psi}^{(j)}(\Gamma^{\mu}(\partial_\mu+\frac{1}{4}\omega_{\nu\lambda,\mu}\Gamma^{\nu\lambda})-m^{(j)})\psi^{(j)} \\
		& S_{ bdy}=\frac{i}{2} \int_{bdy} d^{3}x\;\sqrt{-h} ({\bar\psi}^{(1)}\psi^{(1)}+{\bar\psi}^{(2)}\psi^{(2)})\label{bdryaction},\\
		&S_{g,A,\Phi }= \int d^4x\sqrt{-g} \left(R+\frac{6}{L^2}-\frac{1}{4}F^2_{\mu\nu} +|\mathcal{D}_{\mu}\Phi |^2 -m^{2}_{\Phi}|\Phi |^{2}\right),\\
		& S^A_{int} =\int d^{4}x\; \sum_{i,j=1}^{2}\sqrt{-g}\bar{\psi}^{(i)}(\mathbb{q}^{(ij)} \Gamma^{\mu}A_\mu)\psi^{(j)} \\
		& S^B_{int}=\int d^{4}x\left(  \bar\psi^{(1)}\, B_{xy}\Gamma^{xy}\, \psi^{(2)}+h.c\right),
		\label{eq:action4}
	\end{align}
	$\omega_{\nu\lambda,\mu}$ is the spin connection, $A_\mu=\mu\delta_{\mu t}(1-\frac{z}{z_H})$ and \(\mathbb{q}^{(ij)}\) charge matrix for \(A_\mu\) field in flavor space. The gamma matrices are defined by 
	\begin{align}
		\Gamma^{t}&=\sigma_1 \otimes i \sigma_2, \; \Gamma^{x}=\sigma_1 \otimes \sigma_1,\;
		\Gamma^{y}=\sigma_1 \otimes \sigma_3, \;\Gamma^{z}=\sigma_3 \otimes \sigma_0, \quad\Gamma^{xy}=\mathbb{1} \otimes (-i \sigma_{2}).
	\end{align}
	and geometry is defined by 
	\begin{align}
		ds^2&=-\frac{f(z)}{z^2}dt^2+\frac{dx^2+dy^2}{z^2}+\frac{dz^2}{z^2f(z)},\quad h= {g}{g^{zz}} \\ f(z)&=1-\frac{z^{3}}{z_H^{3}}+\frac{z^4\mu^2}{z_H^2}-\frac{\mu z^3}{z_H}, \quad z_H=\frac{1}{2\pi T+\sqrt{4\pi^2T^2+3\mu^2}},
		\label{eq:geo}
	\end{align}
	
	{\it Source   identification: }
	From  the action (\ref{eq:action1})-(\ref{eq:action4}) above, we can get derivation of total action:
	\begin{equation} \label{deltaS}
		\begin{aligned}
			\delta S & = & \frac{i}{2}\int_{\partial M} d^3x \sum_{j=1}^{2} \sqrt{-h}
			\left(\psi^{(j)\dagger}\Gamma^t(1+\Gamma^z)\delta\psi^{(j)}+\delta\psi^{(j)\dagger}\Gamma^t(1-\Gamma^z)\psi^{(j)}\right). 	
		\end{aligned}
	\end{equation}
	Introducing \(\zeta^{(j)}\) by 
	\be
	\label{wave1}
	\psi^{(j)}= (- g g^{z z})^{- 1 / 4} \zeta^{(j)} e^{- i \omega t + i k_{x} x + i k_{y} y},
	\hbox{ for }  j=1,2, 
	\ee and gamma metric \(\Gamma_\pm= \half(1\pm\Gamma^z)\) we can rewrite (\ref{deltaS}) as 
	\begin{equation}
		\begin{aligned}
			\delta S & = & i \int_{\partial M} d^3x 
			({\bar\zeta}^{(1) }\Gamma_+\delta\zeta^{(1)}+\delta\bar\zeta^{(1) }\Gamma_-\zeta^{(1)}+\bar\zeta^{(2) }\Gamma_+\delta\zeta^{(2)}+\delta\bar\zeta^{(2) }\Gamma_-\zeta^{(2)}) .
		\end{aligned}
	\end{equation}
	From this, we can see that we  need to choose  \((\zeta^{(1)}_1,\zeta^{(1)}_2,\zeta^{(2)}_1,\zeta^{(2)}_2):=\xi^{(S)} \) as the source that make the variation of total action zero. 
	
	Once the source is identified,  we can determine the condensation as the momentum conjugation of the source variable from the total action. Due to the equation of motion, the total boundary action is given by the boundary terms only:
	\begin{align}
		S_{tot} = & \frac{i}{2}\int_{bdy} d^3x \sqrt{-h} (\bar{\psi}^{(1)}\psi^{(1)}+\bar{\psi}^{(2)}\psi^{(2)}) \\
		=&\frac{1}{2}\int_{bdy} d^3x (\zeta^{(1)\dagger}\Gamma^ti\zeta^{(1)}+\zeta^{(2)\dagger}\Gamma^ti\zeta^{(2)}) \nonumber\\
		=&\frac{1}{2}\int_{bdy} d^3x \sum_{j=1}^{2} \left(i\zeta^{(j)*}_1\zeta^{(j)}_4-i\zeta^{(j)*}_2\zeta^{(j)}_3+i\zeta^{(j)*}_3\zeta^{(j)}_2-i\zeta^{(j)*}_4\zeta^{(j)}_1\right),
		\label{eq:totbdy}
	\end{align}
	from which we can identify 
	\((\zeta^{(1)}_3,\zeta^{(1)}_4,\zeta^{(2)}_3,\zeta^{(2)}_4):=\xi^{(C)}\) .\\
	
	{\it Dirac equations } are given by   
	\be
	\begin{aligned}
		(\Gamma^{M}\partial_M+\frac{1}{4}\omega_{\nu\lambda,M}\Gamma^{\nu\lambda}-m^{(1)}+\mathbb{q}^{(11)}A_t \Gamma^{t})\psi^{(1)}+ (B_{xy}\Gamma^{xy}+\mathbb{q}^{(12)}A_t \Gamma^{t})\psi^{(2)}=0,	\\
		(\Gamma^{M}\partial_M+\frac{1}{4}\omega_{\nu\lambda,M}\Gamma^{\nu\lambda}-m^{(2)}+\mathbb{q}^{(22)}A_t \Gamma^{t})\psi^{(2)}+ (B_{xy}\Gamma^{xy}+\mathbb{q}^{(21)}A_t \Gamma^{t})\psi^{(1)}=0.
	\end{aligned}
	\ee
	
	In terms of $\zeta$'s,   the equations are given by 
	\begin{align*}
		\left(\Gamma^{z}\partial_z-\frac{i}{f(z)}(\omega+\mathbb{q}^{(11)}) A_t \Gamma^{t}-\frac{i}{\sqrt{f(z)}}(k_x\Gamma^x+k_y\Gamma^y+\frac{m^{(1)}}{z})\right)&\zeta^{(1)} \\ + \left(B_{xy}\Gamma^{xy}-\frac{i}{f(z)}\mathbb{q}^{(12)}A_t \Gamma^{t}\right) & \zeta^{(2)} =0,	\\
		\left(\Gamma^{z}\partial_z-\frac{i}{f(z)}(\omega+\mathbb{q}^{(22)}) A_t \Gamma^{t}-\frac{i}{\sqrt{f(z)}}(k_x\Gamma^x+k_y\Gamma^y+\frac{m^{(2)}}{z})\right)&\zeta^{(2)} \\+ \left(B_{xy}\Gamma^{xy}-\frac{i}{f(z)}\mathbb{q}^{(21)}A_t \Gamma^{t}\right)&\zeta^{(1)}=0.
	\end{align*}
	
	Now, we can re-write the equation of motion  in terms of 
	\(\xi^{(S)}\), \(\xi^{(C)}\): 
	\begin{align}	
		\mathbb{M}_{11}\xi^{(S)}+\mathbb{M}_{12}\xi^{(C)}+\partial_{z}\xi^{(S)}=0\label{eq:eom1}, \\
		\mathbb{M}_{21}\xi^{(C)}+\mathbb{M}_{22}\xi^{(S)}+\partial_{z}\xi^{(C)}=0. \label{eq:eom2}
	\end{align}
	where
	\begin{gather*}
		\mathbb{M}_{11}=-\mathbb{M}_{21}=-\frac{i}{z\sqrt{f(z)}}
		\begin{pmatrix}
			m^{(1)}\sigma_0 & \mathbb{Q} \\
			\mathbb{Q} & m^{(2)}\sigma_0 
		\end{pmatrix}, \quad
		\mathbb{M}_{12}=-\mathbb{M}_{22}=
		\begin{pmatrix}
			\mathbb{N}(\mathbb{q}^{(11)}) & \mathbb{P}(\mathbb{q}^{(12)}) \\
			\mathbb{P}(\mathbb{q}^{(21)}) & -\mathbb{N}(\mathbb{q}^{(22)})
		\end{pmatrix}, \quad \\
		\mathbb{N}(q)=\frac{i}{\sqrt{f(z)}}
		\begin{pmatrix}
			k_y & -\frac{(\omega+q A_t)}{\sqrt{f(z)}}+k_x \\
			\frac{(\omega+q A_t)}{\sqrt{f(z)}}+k_x & -k_y
		\end{pmatrix}, \; 
		\mathbb{P}(q)=
		\begin{pmatrix}
			0 & -i\frac{q A_t}{{f(z)}} \\
			i\frac{q A_t}{{f(z)}} & 0
		\end{pmatrix}, \; 
		\mathbb{Q}=
		\begin{pmatrix}
			0 & iB_{xy} \\
			-iB_{xy} & 0
		\end{pmatrix},
	\end{gather*}
	where \(\sigma_0\) is identity matrix. Because $\xi^{(S)}$ and $\xi^{(C)}$  are of 4 components, there are 4-independent solutions. If we denote 4 solutions for $\xi^{(S)}$ by $\xi^{(S,i)}, i=1,\cdots ,4$, then arbitrary $\xi^{(S)}$ can be expressed as 
	\begin{align}
		\xi^{(S)}=\sum_{i=1}^{4} c_{i} \xi^{(S,i)}=\mathbb{S}(z)\mathbf{c},
	\end{align}
	where $ \mathbb{S}(z)$ is the 4 by 4 matrix whose $i$-th column is given by $\xi^{(S,i)}$, and $c$ is a column vector whose $i$-th component is $c_{i}$. Similar expression is available for $\xi^{(C)}$. Furthermore, from (\ref{eq:eom1}) and (\ref{eq:eom2}), $\xi^{(C)}$ can be expressed in terms of $\xi^{(S)}$, therefore we should use the same $c_{i}$ for $\xi^{(C)}$ also. Namely,
	\begin{align}
		\xi^{(S)}=\mathbb{S}(z)\mathbf{c}, \quad \xi^{(C)}=\mathbb{C}(z)\mathbf{c}. \label{eq:mrdef}
	\end{align}
	Substituting these to (\ref{eq:eom1}) and (\ref{eq:eom2}), we find:
	\begin{align}
		\partial_z\mathbb{S}(z)+\mathbb{M}_{11}\mathbb{S}(z)+\mathbb{M}_{12}\mathbb{C}(z)=0, \label{eq:mEQ1} \\
		\partial_z\mathbb{C}(z)+\mathbb{M}_{21}\mathbb{C}(z)+\mathbb{M}_{22}\mathbb{S}(z)=0, \label{eq:mEQ2}
	\end{align}
	because $c$ is an arbitrary vector in the solution space. 
	Now, we consider the near boundary behavior of $\xi^{(S)}$ and $\xi^{(C)}$, which are given by
	\begin{align}
		\zeta^{(j)} = \begin{pmatrix} A_1^{(j)} z^{m}+B_1^{(j)} z^{1-m} \\ A_2^{(j)} z^{m}+B_2^{(j)} z^{1-m} \\ C_1^{(j)} z^{1+m}+D_1^{(j)} z^{-m} \\ C_2^{(j)} z^{1+m}+D_2^{(j)} z^{-m} \end{pmatrix} \quad \hbox{for}\ j=1,2.
		\label{eq:xi}
	\end{align}
	where $A$, $B$, $C$, and $D$ are two-component spinors. If $|m|<1/2$, $ A, D$ terms are leading ones. Therefore 
	\begin{align}
		\zeta^{(j)} \simeq ( A_1^{(j)} z^{m}, A_2^{(j)} z^{m}, D_1^{(j)} z^{-m},  D_2^{(j)} z^{-m}) \quad \hbox{for}\ j=1,2.
		\label{eq:bdyxi}
	\end{align}
	From (\ref{eq:bdyxi}), 
	\begin{align}
		\xi^{(S)}& \simeq ( A_1^{(1)} z^{m}, A_2^{(1)} z^{m}, A_1^{(2)} z^{m}, A_2^{(2)}z^{m}), \\
		\xi^{(C)}& \simeq ( D_1^{(1)} z^{-m},  D_2^{(1)} z^{-m}, D_1^{(2)} z^{-m},  D_2^{(2)} z^{-m}).
	\end{align}
	Therefore, if we define 
	\begin{align}
		U(z)={\rm diag}(z^m,z^m,z^{-m},z^{-m}),
	\end{align}
	then the near boundary behavior of 
	$\xi^{(S)}$ and $\xi^{(C)}$ can be written as
	\begin{align}
		\xi^{(S)}&=\mathbb{S}(z)\mathbf{c} \simeq U(z)\mathbb{S}_0\mathbf{c}, \label{eq:ns} \\
		\xi^{(C)}&=\mathbb{C}(z) \mathbf{c}\simeq U(z)^{-1}\mathbb{C}_0\mathbf{c}, \label{eq:nc} 
	\end{align}
	where $ \mathbb{S}_{0}$ is a matrix whose i-th column is given by the coefficients of the leading terms in $\xi^{(S,i)}$. A similar description works for $\mathbb{C}_{0}$. Defining 
	\begin{align}
		\mathcal{J}=\mathbb{S}_0\mathbf{c}, \quad \mathcal{C}=\mathbb{C}_0\mathbf{c}, \label{SCc}
	\end{align}
	we get
	\begin{align}
		\xi^{(S)}\simeq U(z)\mathcal{J}, \quad \xi^{(C)}\simeq U(z)^{-1} \mathcal{C}. \label{eq:nsc}
	\end{align}	
	It is easy to see that 
	\begin{align}
		\mathcal{J}&=( A_1^{(1)}, A_2^{(1)}, A_1^{(2)}, A_2^{(2)}),\quad \\ \mathcal{C}&=( D_1^{(1)} ,  D_2^{(1)} , D_1^{(2)},  D_2^{(2)}).
	\end{align}
	\\
	{\it Green's function: } Using notation of \(\xi_S,\ \xi_C\) we can rearrange total boundary action (\ref{eq:totbdy}),
	\begin{align}
		\nonumber
		S\mid_{bdy}= & \frac{1}{2}\int_{bdy} d^3x[\xi^{(S)\dagger}(-\sigma_{0}\otimes \sigma_{2})\xi^{(C)}+h.c]&\\ 
		=&\frac{1}{2}\int_{bdy} d^3x\left(\xi^{(S)\dagger}\tilde{\Gamma}\xi^{(C)}+h.c\right). &
	\end{align}
	with \(\tilde{\Gamma}=-\sigma_{0}\otimes \sigma_{2}\). Using (\ref{eq:nsc}), we get
	\begin{align}
		S\mid_{bdy}=&\frac{1}{2}\int_{bdy} d^3x\mathcal{J}^{\dagger}\tilde{\Gamma}\mathcal{M}.
		\label{eq:gr1}
	\end{align}
	For (\ref{SCc}), we can rewrite:
	\begin{align}
		\mathcal{M}=\mathbb{C}_0\mathbb{S}_0^{-1}\mathcal{J},
	\end{align}
	so that (\ref{eq:gr1}) becomes 
	\begin{align}
		\centering
		\nonumber
		\frac{1}{2}\int_{bdy} d^3x\mathcal{J}^{\dagger}\tilde{\Gamma}\mathcal{M}=&\frac{1}{2}\int_{bdy} d^3x\mathcal{J}^{\dagger}\tilde{\Gamma}\mathbb{C}_0\mathbf{c}\\ \nonumber
		=&\frac{1}{2}\int_{bdy} d^3x\mathcal{J}^{\dagger}\tilde{\Gamma}\mathbb{C}_0\mathbb{S}_0^{-1}\mathcal{J}\\ 
		=&\frac{1}{2}\int_{bdy} d^3x\mathcal{J}^{\dagger}\mathbb{G}_0 \mathcal{J}.
	\end{align}
	Finally, we can define \(\mathbb{G}_0=\tilde{\Gamma}\mathbb{C}_0\mathbb{S}_0^{-1}\) and this is the Green function of the system.\\
	
	{\it Spectral function by flow equation: } Combining the above equations \eqref{eq:mEQ1},\eqref{eq:mEQ2} and using the definition of green's function,  the flow equations  become 
	\be
	\begin{aligned}
		\tilde{\Gamma} \mathbb{M}_{21} \tilde{\Gamma} \mathbb{G}(z)+ \tilde{\Gamma} \mathbb{M}_{22}-\mathbb{G}(z) \mathbb{M}_{11}-\mathbb{G}(z) \mathbb{M}_{12}\tilde{\Gamma} \mathbb{G}(z)+\partial_{z}\mathbb{G}(z)=0.
	\end{aligned}
	\ee
	We want to express the boundary Green function $\mathbb{G}_0 $ in terms of the bulk quantity  $\mathbb{G}(z)$ near the AdS boundary. From the boundary  behaviors of $\zeta^{(1)},\zeta^{(2)}$  given in (\ref{eq:bdyxi}), we can express those of $\xi^{(S)}$ and $\xi^{(C)}$:
	If we substitute the expression \eqref{eq:nc} to the definition of the Green function,  
	\begin{align}
		\mathbb{G}(z)&=\tilde{\Gamma}\mathbb{C}(z)\mathbb{S}(z)^{-1} \nonumber \\
		&\simeq \tilde{\Gamma}U(z)\mathbb{C}_0\mathbb{S}_0^{-1}U(z)^{-1} \nonumber \\
		&=U(z)\mathbb{G}_0U(z)^{-1}.
	\end{align}
	In the third line, we use the ${\tilde{\Gamma}}^{2}=\mathbb{1}_{4\times4}$ and $\tilde\Gamma U(z)\tilde\Gamma=U(z)$. Therefore the boundary Green function $\mathbb{G}_0$ is given as follows:
	\begin{align}
		\mathbb{G}_0=\lim_{z\rightarrow0}U(z)^{-1}\mathbb{G}(z)U(z). \label{eq:dobg}
	\end{align} 
	Solving this flow equation, with boundary condition, we can get spectral function:
	\begin{align}
		A(k,w)=\Im [Tr[\mathbb{G}_0]].
	\end{align}
	
	{\it Horizon   value of the Green Function: }
	Here we motivate the use of the flow equation by showing the regularity of the $\mathbb{G}(z)$ at the horizon. In this section, we derived the near horizon behavior of the fermion, which can be recast as
	\begin{align}
		\zeta^{(j)}=\begin{cases}
			(1-z/z_H)^{-\frac{i \omega z_H}{3}}(A_1^{(j)},A_2^{(j)},-A_2^{(j)},A_1^{(j)})^T , & \hbox{ for the infalling},\\
			(1-z/z_H)^{\frac{i \omega z_H}{3}} (A_1^{(j)},A_2^{(j)},A_2^{(j)},-A_1^{(j)})^T, & \hbox{ for the outgoing},
		\end{cases} 
	\end{align}
	for \(j=1,2\). We choose an infalling condition for each flavor. Then we can construct the horizon behavior of $\xi^{(S)}$ and $\xi^{(C)}$:
	\begin{align}
		\xi^{(S)}&=Z(A_1^{(1)},A_2^{(1)},A_1^{(2)},A_2^{(2)})^T, \\
		\xi^{(C)}&=Z(-A_2^{(1)},A_1^{(1)},-A_2^{(2)},A_1^{(2)})^T,	
	\end{align}
	where, $Z=(1-z/z_H)^{-\frac{i w z_H}{3}}$. Using the matrix representation, for appropriate \(A_i^{(j)}\),\(B_i^{(j)}\) with \(j=1,2\)
	\begin{align}
		\mathbb{S}(z)\simeq Z
		\begin{pmatrix}
			1 & 1 & 1 & 1 \\
			1 & -1 & 1 & 1 \\
			1 & 1 & -1 & 1 \\
			1 & 1 & 1 & -1 \\
		\end{pmatrix} , \qquad
		\mathbb{C}(z)\simeq Z
		\begin{pmatrix}
			-1 & 1 & -1 & -1 \\
			1 & -1 & 1 & 1 \\
			-1 & -1 & -1 & 1 \\
			1 & 1 & -1 & 1 \\
		\end{pmatrix}.
	\end{align}
	The horizon value of the matrix Green function is given by
	\begin{align}
		\mathbb{G}(z)=\tilde{\Gamma}\mathbb{C}(z)\mathbb{S}^{-1}(z)=i \mathbb{1}_{4\times4},
	\end{align}
	which is rather surprising: $\mathbb{G}(z)$ is constant near the horizon while $\mathbb{S}(z)$ and $\mathbb{C}(z)$ are singular at $z=z_{H}$. This is very important in the numerical calculation, which is why we want to use flow equations.
	
	\subsection{Holographic Flat band over a Disk in momentum space}
	
	For  $d=3, p=2, \Delta=2[\psi]=2$, in $AdS_{4}$ can set $m_\Phi^2=0$ and  asymptotic form  of $B_{xy}$
	is given by   \be
	B_{xy}=B_{xy}^{(-1)} z^{-1} +B^{(0)}_{xy}. 
	\ee
	\begin{figure}[H]
		\centering
		\subfigure[Band structure for \(t_{\perp}=1\)]
		{\includegraphics[width=5cm]{./fig/ABC20.png}}
		\subfigure[Spectrum for \(B_{xy}^{(-1)}=1\)]
		{\includegraphics[width=5cm]{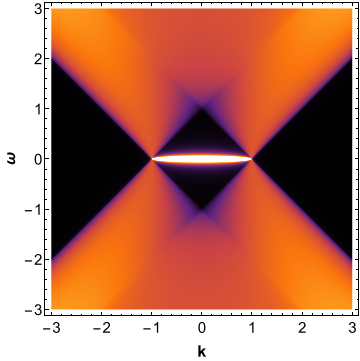}}
		\subfigure[\(\omega=0\) slice of spectrum]
		{\includegraphics[width=5cm]{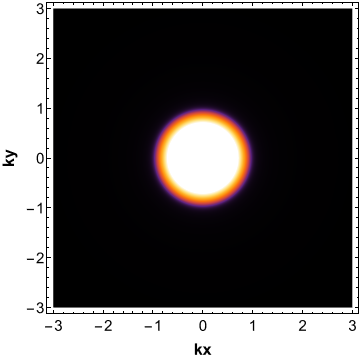}}
		\caption{(a) Band structure of ABC-MLG by tight-binding Model, (b) holographic Spectral function of 2 flavor fermion with \(B_{xy}\) interaction with \(B_{xy}^{(-1)}=1\), \(k_y=0\) (c) \(\omega\) slice of holographic spectral function}
		\label{fig:compare}
	\end{figure}
	In Fig. \ref{fig:compare} (b) and (c), we draw the spectral function with \(B^{(-1)}_{xy}=2\), \(B^{(0)}_{xy}=0\).  
	We can see the presence of the disk-shaped flat band. In this case, the size of the flat band can be identified as the disk's radius, which depends only on the strength of \(B_{xy}\). 
	The following relationship holds.   
	\be
	R_{holo}= B_{xy}^{(-1)}.
	\label{eq:holofbr}
	\ee
	
	\subsection{Flat band parameter}
	
	So far, we have considered two models, the Holographic and tight binding models,  realizing the flat bands and analyzing the parameter dependence of the size of the flat band in each model. 
	We find that the holographic model describes the
	limit of large stacking numbers. 
	By identifying $R_{tb}$ and $R_{holo}$, we can determine the relationships between the parameters in the two models.   
	From (\ref{eq:tbfbr}) and (\ref{eq:holofbr}),  we get
	\begin{equation}
		B^{(-1)}_{xy}=t_{\perp} .
	\end{equation} 
	In Fig.\ref{fig:compare} (b) and (c), we compared the spectrum of the two models, which shows striking similarity. 
	We learned that the interaction term describes the ABC-stacking method for the tight-binding model, which is matrix  \(\Gamma_{xy}\) in the holographic model. In the tight-binding model, the flat band structure is created by the inter-layer interaction of the ABC stacked graphene system. On the other hand,  in the holographic model, the spectrum is determined by \(\Gamma_{xy}\sim  \half [\Gamma^{x},\Gamma^{y}]\) which is the generator of the rotation in the x-y plane for the spinors. Indeed, the spectral function of both models is rotationally symmetric.
	\footnote{It turns out that 
		if we perform the calculation in  AdS$_{5}$, the flat band is localized in $k_{z}=0$ plane in k-space so that the translational symmetry is broken even though we did not introduce the boundary of the boundary. 
		On the other hand,  in the tight-binding model, which describes non-interacting material with a natural boundary, the translational symmetry in the z-direction is broken by the very existence of the boundary, and the flat band is nothing but the surface mode, as it was shown explicitly in section 2. 
		One surprising consequence is that the dispersion relation defines the dimensions of the geometry as equal in both AdS$_{4}$ and AdS$_{5}$. 
		In this way, the AdS$_{5}$ theory also describes the 
		multi-stacking of two-dimensional layers.  
	}

	\section{Bending in flat band system} \label{sec:chemical}
	
	\begin{figure}[H]
		\centering
		\subfigure[ARPES data \cite{PhysRevB.97.245421}]
		{\includegraphics[width=5cm]{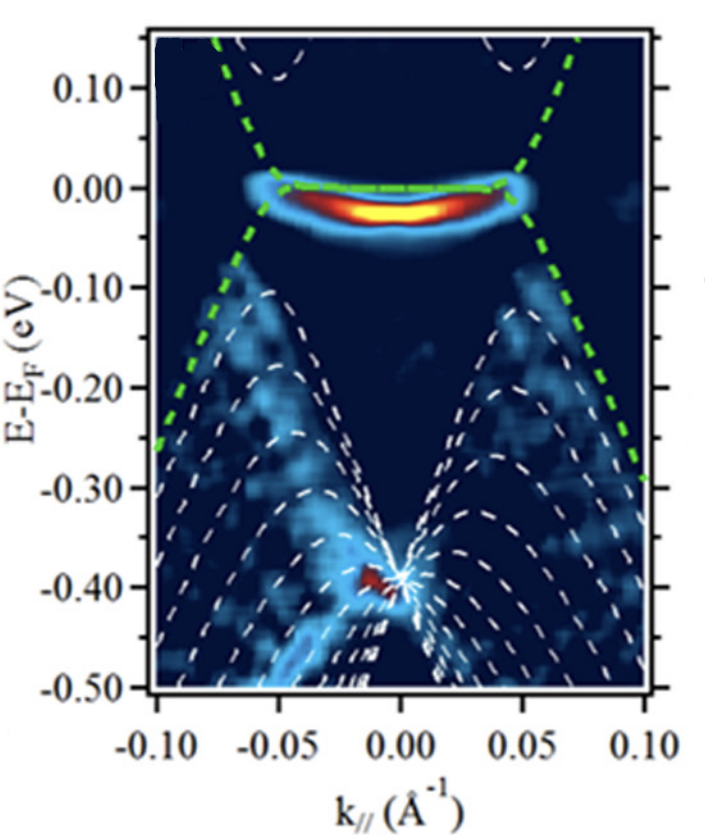}}
		\subfigure[Holographic model]
		{\includegraphics[width=4.7cm]{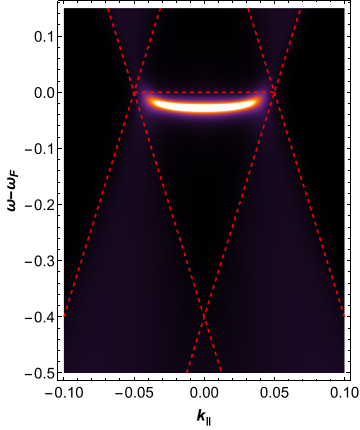}}
		\caption{Comparision experimental data and holographic  model: (a) Angle-resolved photoemission spectroscopy(ARPES) data of ABC-MLG film for 14-layers, dashed line denote band structure especially green one means surface state \cite{PhysRevB.97.245421}, (b) the spectral function of holographic ABC-MLG model with \(B^{(-1)}_{xy}=0.05\), \(\mu=0.075\), and some scaling for \(\omega\), red dashed line denote overall bend structure coming from the tight-binding model.}
		\label{mu}
	\end{figure}
	In real systems, it is known that the flat band has instability. Therefore there is inevitably a slight deformation in the flat band system. Band bending is the easiest deformation. In fact, we can easily see this in a real experiment. See in Fig. \ref{mu} (a), which shows the bending of the flat band. 
	This bending effect in the ABC-MLG  can have two interpretations for its origin. 
	The first is due to the chemical potential, and the second is due to
	the  spin polarization \cite{PhysRevB.97.245421,PhysRevB.95.075422}.
	While it is difficult to reproduce the bending in the tight-binding model, 
	in holography, we can reproduce the bending of the flat band by considering the chemical potential of a  \(U(1)\) field. Notice that in the tight-binding model, the chemical potential just shifts the overall spectrum. 
	The effect of the chemical potential in holography is described by the coupling of the \(U(1)\) field 
	\be
	A_t= \mu(1-\frac{z}{z_H}) .
	\ee
	to the fermion. 
	The first term shifts the entire spectrum, whereas the second term
	gives a density effect, including mutual repulsion, which can cause bending in the flat band system. See Fig \ref{mu}.
	
	However, we remark that the description of the bending by the U(1) gauge field does not eliminate the spin effect as the origin of the bending. Because when we include the electric field, the magnetic field enters as the $B=\vec{v}\times \vec{r} \frac{dV}{dr}$ so that $H_{int}=-\vec{\mu}\cdot \vec{B} \sim S\cdot L\frac{dV}{dr}$. Such magnetic field  act as an effective  spin chemical potential
	which has different signs for the different spin states (up or down).   
	Notice that in the Dirac equation,  the spin-orbit coupling is always included. 

	Also, the pseudo spin structure is naturally contained when there is a sublattice interaction. That is, if we consider spin polarization,  interlayer coupling \(t_{\perp}c^{\dagger}_{B_{j\uparrow}}c_{A_{j+1\downarrow}}\) appears. This model does not distinguish the layers' positions in the \(z\) direction. So, \(B_{j\uparrow}\) and \(A_{j+1\downarrow}\), which belong in the different layers,  are considered to be in the same position. See Fig \ref{fig:stackingstruc}. So, the hopping term \(t_{\perp}c^{\dagger}_{B_{j\uparrow}}c_{A_{j+1\downarrow}}\) can   be considered as the on-site energy that gives opposite energy to  A and B sublattices.
	
	We can see the polarization effect as an analog of the chemical potential \(\mu\) in the \(U(1)\) field. In the ABC-MLG system, the flat band is from two band origins from the top and bottom layers. These surface states are ferrimagnetic and anti-ferromagnetically coupled to each other. And they have opposite spin for each sublattice \(A\), \(B\) \cite{lee2014competition, PhysRevB.86.115447}. 
	In our model, we have two flavors, and their index   is interpreted as sublattice index in the ABC-MLG system:  when \(A_t\) field comes with interaction term $
	\bar{\psi}^{(i)}(\mathbb{q}^{(ij)} \Gamma^{\mu}A_\mu)\psi^{(j)}$ 
	for off-diagonal charge with 
	$\mathbb{q}=\begin{pmatrix}
		0 & 1  \\
		1 & 0  \\
	\end{pmatrix}$, after diagonalization, we can give spin polarization effect with charge $\mathbb{q}=\begin{pmatrix}
		1 & 0  \\
		0 & -1  \\
	\end{pmatrix}.$
	And this makes bending in a different direction for each sublattice. In this sense, we can say diagonalization in flavor space makes pseudo-spin.
	\begin{figure}[H]
		\centering
		\subfigure[\(\mathbb{q}^{(ij)}=0\) for \(i,j\in{1,2}\)
		]
		{\includegraphics[width=5cm]{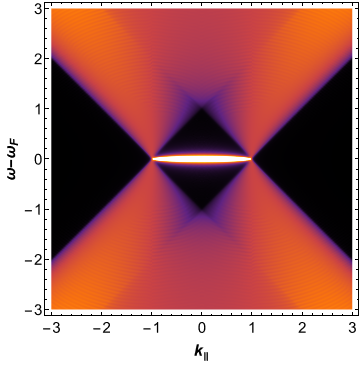}}
		\subfigure[\(\mathbb{q}^{(11)}=\mathbb{q}^{(22)}=1\) charge]
		{\includegraphics[width=5cm]{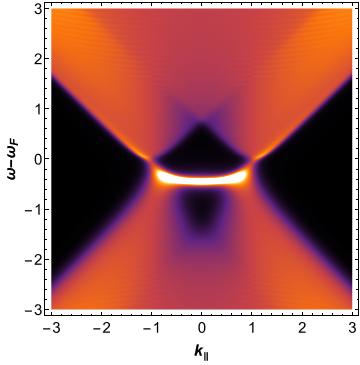}}
		\subfigure[\(\mathbb{q}^{(12)}=\mathbb{q}^{(21)}=1\) charge]
		{\includegraphics[width=5cm]{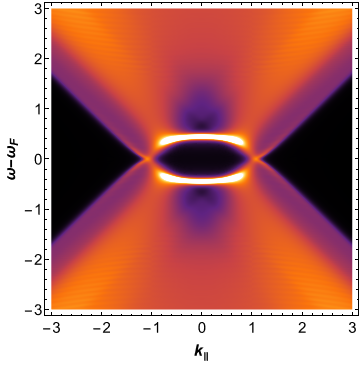}}
		\caption{Bending effect of Holographic ABC-MLG by chemical potential \(\mu\) in different charge matrix with \(B_{xy}^{(-1)}=1\), \(\mu=0.5\)}
		\label{fig:spin}
	\end{figure}
	Fig. \ref{fig:spin} shows two kinds of bending effects in holographic ABC-MLG. We can make design the interaction term that can reproduce the result of ARPES data by choosing the charge matrix $\mathbb{q}$.
	
	\section{Conclusion}
	In this paper, we considered the continuum limit of the tight-binding model of ABC-MLG. The model preserves the rotation symmetry and the scale \(t_{\perp}\) coming from the coupling of two sublattices, $A$ and $B$. As the number of layers goes infinite, the spectrum near the K point has a flat band. 
	Such features can be met by a holographic model where we introduce two flavors \(\psi^{(1)}\) and \(\psi^{(2)}\) that represent sublattices \(A, B\). 
	The interaction term $$\int_{bulk} {\bar\psi^{(1)}}B_{xy}\Gamma^{xy}\psi^{(2)},$$ was introduced to match the spectral shape and the symmetry of the tight binding model.  
	It turns out that the scale of the holographic order parameter \(B_{xy}\)  and the scale of the tight binding model  \(t_{\perp}\) play precisely the same role, and we identify them. 
	In this sense, we can say that we created a low-energy effective model of large \(N\) ABC-MLG in holography. 
	
	Unlike the tight-binding model, in holography, when a chemical potential is applied by \(U(1)\) field, not only the energy shift but also the density effect are realized through the flat band's bending. 
	Table \ref{tab:correspondence}. summarizes the correspondence of the two models we studied.  \begin{table}[H]
		\begin{center}
			\begin{tabular}{ |c|ccc|  }
				\hline
				& Tight binding model & \(\iff\) & Holography \\ \hline
				Spacetime & Semi - 2+1 dim  & \(\iff\) & AdS4 \\  \hline
				Sublattice & A, B  & \(\iff\) & \(\psi^{(1)}, \psi^{(2)}\)
				\\  \hline
				Symmetry & Rotation symmetry & \(\iff\) & 
				Rotation symmetry\\ \hline
				Scale & \(t_{\perp}\) & \(\iff\) & \(B_{xy}\) \\ \hline
			\end{tabular}
			\caption{Correspondence of two models}
			\label{tab:correspondence}
		\end{center}
	\end{table}
	
	\appendix
	
	\section{Reduced hamiltonian}
	\begin{figure}[H]
		\centering
		\begin{tikzpicture}
			\draw[line width=2pt,color=black] (-3, 2) -- (-1.5, 2);
			\draw[line width=1pt,color=gray] (-1.5, 2) -- (-1.5, 0.5);
			\draw[line width=2pt,color=black] (-1.5, 0.5) -- (0, 0.5);
			\node(dots) at (0,0){\vdots};
			\draw[line width=2pt,color=black] (0, -0.5) -- (1.5, -0.5);
			\draw[line width=1pt,color=gray] (1.5, -0.5) -- (1.5, -2);
			\draw[line width=2pt,color=black] (1.5, -2) -- (3, -2);
			\begin{scope}[ebo unit=mm,red,line width=1.5pt]
				\draw [ebo corners=5] (-2.5,-2.5) ++(0,0) rectangle ++(5,5);
			\end{scope}	
			\draw [fill=white](-3,2) circle (0.1);
			\draw [fill=black](-1.5,2) circle (0.1);
			\draw [fill=white](-1.5,0.5) circle (0.1);
			\draw [fill=black](1.5,-0.5) circle (0.1);
			\draw [fill=white](1.5,-2) circle (0.1);
			\draw [fill=black](3,-2) circle (0.1);
			\node[right] at (4,2) {Top layer};
			\node[left] at (-3,2) {\(A_1\)};
			\node[right] at (-1.5,2) {\(B_1\)};
			\node[below] at (-1.5, 0.4) {\(A_2\)};
			\node[above] at (1.5, -0.4) {\(B_{N-1}\)};
			\node[right] at (4, 0) {Bulk};
			\node[above, red] at (0, 2.5) {Effective coupling};
			\node[left] at (1.5, -2) {\(A_N\)};
			\node[right] at (3, -2) {\(B_N\)};
			\node[right] at (4, -2) {Bottom layer};	
		\end{tikzpicture}
		\caption{Semi-effrctive atomic structure of ABC-MLG. The red box is the effective coupling between the top and bottom layers.}
		\label{effective}
	\end{figure}
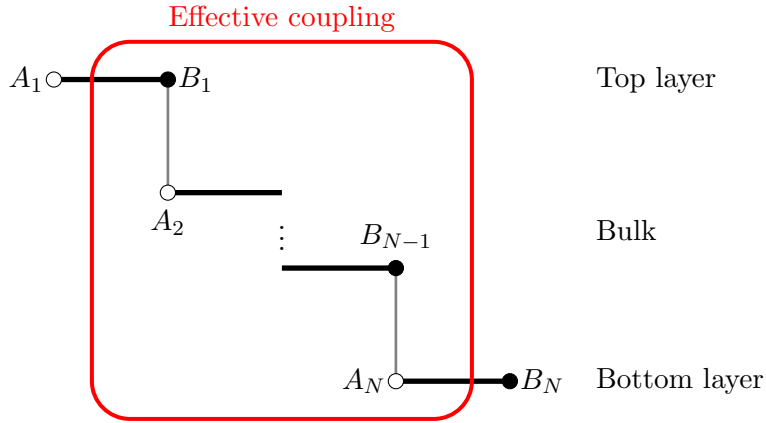
	To see the parameter dependence of the flat band system, we consider the low-energy effective Hamiltonian is reduced to the subspace with the basis \(\left| A_1 \right\rangle, \left| B_N \right\rangle\) \cite{PhysRevLett.96.086805, PhysRevB.75.155424, PhysRevB.77.155416}.
	 
	We can reorder the wave function
	\((A_1,B_1,A_2,B_2,\cdots ,A_{N-1},B_{N-1},A_N,B_N)\) to \\
	\((A_1,B_{N},A_{2},\) \(B_3,\cdots ,A_{N-1},B_2,A_{N},B_1)\). 
	\(N=3\) with   \(U_j=0\) case, 
	we can rewrite (\ref{eq:fullH}) as
	\begin{align}
		\mathcal{H}(k)=\begin{pmatrix}  0 & 0 &0 &0 &0 &vp_{-}  \\  0 & 0 &0 &0 &vp_{+} &0   \\  0 & 0 &0&vp_{-}&0&t_{\perp}  \\ 0& 0 & vp_{+} &0&t_{\perp}&0\\  0 & vp_{-} &0&t_{\perp}&0&0\\  vp_{+} & 0 &t_{\perp}&0&0&0\\\end{pmatrix} \equiv \begin{pmatrix} H_{11} & H_{12} \\ H_{21} & H_{22} \end{pmatrix},
	\end{align}
	with block matrices
	\begin{align*}
		H_{11}=\begin{pmatrix} 0 & 0  \\ 0 & 0   \end{pmatrix}, && H_{12}=\begin{pmatrix} 0 & 0 & 0 & vp_{-} \\ 0 & 0 & vp_{+} & 0 \end{pmatrix}, \\
		\\
		H_{21}=\begin{pmatrix} 0 & 0 \\ 0 & 0 \\ 0 & vp_{-} \\ vp_{+} & 0  \end{pmatrix}, && H_{22}=\begin{pmatrix} 0 & vp_{-} & 0 & t_{\perp} \\ vp_{+} & 0 & t_{\perp} & 0 \\ 0 & t_{\perp} & 0 & 0  \\t_{\perp} & 0 & 0 & 0 \end{pmatrix}.
	\end{align*}
	In terms of these block matrices, we have the identity
	\begin{align}
		\det(\mathcal{H}-E)=\det(H_{11}-H_{12}(H_{22}-E)^{-1}H_{21}-E)\det(H_{22}-E).
	\end{align}
	In $E \ll t_{\perp}$ limit, we can replace $H_{22}-E$  by $ H_{22}$. Then   we can reduce hamiltonian as
	\begin{align}
		\mathcal{H}^{eff}_{N=3}\equiv H_{11}-H_{12}H_{22}^{-1}H_{21}=-{1\over t_{\perp}^2} \begin{pmatrix} 0 & (vp_{-})^{3} \\ (vp_{+})^{3} & 0 \end{pmatrix}.
	\end{align}
	For the general $N$ and with the presence of \(U_j\), it can be generalized as \cite{PhysRevB.81.125304}.
	\begin{align}
		\mathcal{H}^{eff}=\begin{pmatrix} U_1 & t_{\perp}(vp_{-} / t_{\perp})^{N} \\  t_{\perp}(vp_{+} / t_{\perp})^{N} & U_N \end{pmatrix} .
	\end{align}
	For convenience, we set \(U_1+U_N=0\). Then the eigenvalues  of effective hamiltonian are given by
	\begin{equation}
		\varepsilon^{eff}=\pm\sqrt{ t_{\perp}^2(vp/ t_{\perp})^{2N}+(\Delta U/2)^2},
		\label{effspec}
	\end{equation}
	where \(\Delta U=U_1-U_N\). This shows the lowest energy band of ABC-MLG, and the size of the flat band depends on the stacking number \(N\) with scale \(t_{\perp}\). Here \(\Delta U\) plays a role in chemical potential, and it sifts the lowest energy modes.
	If we replace (\ref{eq:geo}) as Lishitz geometry:
	\begin{align}
		ds^2&=\frac{-f(z)}{z^{2N}}dt^2+\frac{dx^2+dy^2}{z^2}+\frac{dz^2}{z^2f(z)}. \quad 
	\end{align}
	We can get dispersion relation:
	\begin{align}
		\omega \sim k^{N}.
	\end{align}
	For \(N=20\) non-interacting Lifshitz fermion, we can draw spectral functions like:
	\begin{figure}[H]
		\centering
		\subfigure[\(N=20\)]
		{\includegraphics[width=5cm]{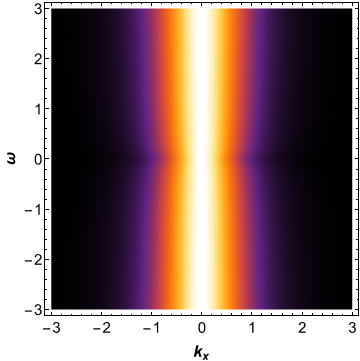}}
		\subfigure[\(N=20\)]
		{\includegraphics[width=5cm]{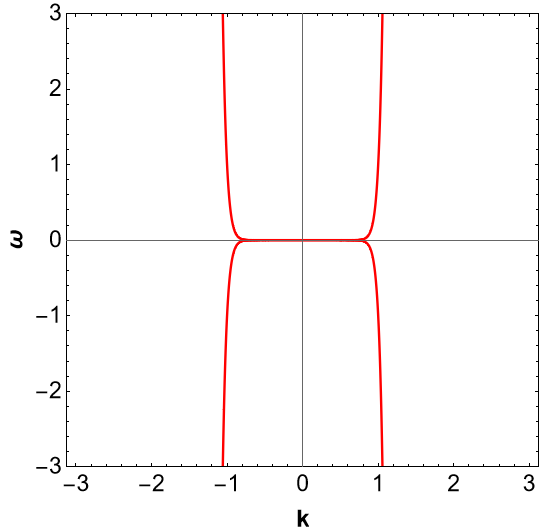}}
		\caption{(a) Spectral function of non-interacting fermion for several Lishitz geometries (b) Band structure of ABC-MLG with several layer stacking order \(N\)'s with \(t_{\perp}=1\) and \(U_{j}=0\) by reduced hamiltonian.}
		\label{fig:AMGlishitz}
	\end{figure}
	This spectral function is the same as that obtained using reduced hamiltonian (\ref{effspec}), which is a low-energy effective model of ABC-MLG. We can say that the tight-binding model's low-energy effective hamiltonian can be described by holography Lifshitz theory.
	
	\acknowledgments
	This work is supported by Mid-career Researcher Program through the National Research Foundation of Korea grant No. NRF-2021R1A2B5B02002603. 
	We also thank the APCTP for the hospitality during the focus program, “Quantum Matter and Entanglement with Holography”, where part of this work was discussed.
	
	\bibliographystyle{JHEP}
	\bibliography{Refs_ABC-MLG.bib}
\end{document}